\documentclass{article}                     

\usepackage{graphicx}
\usepackage{bm}
\usepackage{amsmath}
\usepackage{amsfonts}
\begin{document}

\title{Possible connections between relativity theory and a version of quantum theory based upon theoretical variables}

\author{Inge S. Helland \\
  Department of Mathematics, University of Oslo \\P.O. Box 1053 Blindern, N-0316 Oslo, Norway\\
}

\date{}

\maketitle

\begin{abstract}
An alternative approach towards quantum theory is described, and tentative attempts to connect his approach to special and general relativity are discussed. Important concepts are gauge groups and information/entropy connected to some physical systems. Some recent results on information in connection to black holes are touched upon, and it is indicated how expected information can be argued to be conserved. This argument only depends on what happens outside the black hole. Everything connected to the interior of the black hole is inaccessible.

\end{abstract}

\section{Introduction}
\label{intro}

To find a conceptual basis from which quantum theory and general relativity both can be understood is one of the most challenging problems in modern physics. Many researchers and several research groups have made their proposals on how to attack this problem. The most well-known approaches are the following three: 1) Quantum loop theory. (For a popular partial account, see Rovelli [1]). 2) String theory. (For a brief introduction, see Susskind and Lindsay [2]). 3) The pure mathematical modelling approach. (See for instance Laudal [3]). From my point of view the operational approach by Hardy [4] may be particularly enlightening. Several relevant references can be found in the latter paper.

In contrast to these references, I will rely on a new and different approach towards the axioms of quantum theory. The approach started with the book [5], and has now been further developed in a series of articles. A summary of the theory is now given in [6]. Central to the theory is a simple model of the mind of an observer, a model which may be generalized to the mind of any person. It first relies on what I call theoretical variables, which may be physical variables, but in the process of planning, doing, or interpreting experiments, the variables are also assumed to exist in the mind of a relevant actor. From a mathematical point of view, my only requirement for the theoretical variables is the following: If $\lambda$ is a theoretical variable and $\theta=f(\lambda)$ for some fixed function $f$, then $\theta$ is also a theoretical variable.

Some theoretical variables may be accessible, that is, by experiment or measurement, it is possible at a certain time in the future to obtain as good information about them as we want to. This definition may be unclear to some readers, but again, from a mathematical point of view, I will stress taht the only property that I require about my accessible variables is: If $\lambda$ is accessible, and $\theta=f(\lambda)$ for some fixed function $f$, then $\theta$ is accessible. From a physical point of view, two examples of accessible variables are the theoretical position or theoretical momentum of a particle. I say theoretical since I model measurement as a theoretical value plus random error. This is in the tradition of statisticians, who will regard my theoretical variables as parameters. I have deliberately avoided the word parameter in my theory since this word also has different meanings for a physicist.
 
Another, and perhaps simpler, physical example of an accessible variable, is the spin component in a fixed direction $a$ of some particle with spin. In quantum theory, this is a discrete variable, and for discrete variables, exact values can be obtained by good experiments, say, a Stern-Gerlach experiment.

Now to my model of the mind of some actor: I assume that in some fixed context he has several accessible variables in his mind, say $\theta, \eta, \xi,...$. These may be physical variables as above, but they may also be completely different theoretical variables. \emph{As a first model assumption assume that there exists an inaccessible variable $\phi$ such that each accessible variable is a function of $\phi$.}

In this model assumption, the theoretical variables, in particular the inaccessible variable $\phi$, are seen as \emph{mathematical} variables.

In the two physical examples above, it is easy to give concrete realisations of such a $\phi$. In the first example it can be taken to be the vector (theoretical position, theoretical momentum), which is inaccessible by Heisenberg's inequality. In the second example we can use the abstract spin vector as $\phi$. For an electron, say, the component in direction $a$ can be taken as $\theta^a = f^a(\phi)= \mathrm{sign} (\mathrm{cos} (\phi,a))$. If $\phi$ is given some natural probability distribution, each $\theta^a$ will get the correct distribution.

In general, the existence of $\phi$ must just be seen as a model, but it turns out to be a useful model. In this model, all variables are seen as mathematical variables.

In both the examples above, the relevant, accessible variables mentioned may be seen as maximal, and no `larger' accessible variables may be found by the following partial ordering: Say that $\theta$ is `less than or equal to' $\lambda$ if $\theta=f(\lambda)$ for some function $f$. By using Zorn's lemma on this partial ordering, it follows from the model that maximal accessible variables always exist. These turn out to be important.

This model is developed in two books [5] and [7], and in the papers [8], [9] and [10].

In [6], the model is generalized. Instead of looking upon theoretical variables as \emph{mathematical} variables, I then only concentrate on \emph{physical} variables, and they are all accessible. From a physical point of view, it is not a reasonable  assumption that the inaccessible theoretical variables are assumed to have sharp values. For instance (position, momentum) cannot, taking into account Heisenberg's inequality, be assumed to take physical values $(x,p)$. 

In this generalization, some simple elements of category theory are used. The space $\Omega_\phi$ over which the maximal accessible variable $\phi$ was assumed to vary is replaced by an object $\Omega$, called the space of \emph{notions} for the observer, and morphisms between objects are considered. In particular, I consider a space $K$ of automorphisms from $\Omega$ onto $\Omega$, and the main model assumption above that all accessible variables are functions of $\phi$, is replaced by the following assumption: \emph{For every accessible variable $\eta$ there is a morphism from $\Omega$ onto the space $\Omega_\eta$.}

The notions of the basic space $\Omega$ may be thought of as anything that is in the mind of the given observer. The basic model can be seen in relation to the philosophy of Convivial Solipsism founded by Herv\'{e} Zwirn [11]: Every description of the world must be with respect to the mind of some observer. But different observers may communicate.

In Section 2 below, I show that, by adding suitable symmetry assumptions to this model, essential elements of quantum mechanics emerge. In the theory that is developed here, we may think of theoretical variables and the notions in the mind of some single person. But alternatively, we may also think of theoretical variables and notions in the joint minds of a communicating group of persons. The only difference in the latter case, is that the accessible variables must always be defined in words, in order that communication shall be possible.

In general, one may take the standpoint that every scientific theory (including what I will present below) is coupled to the mind of at least one person or to the joint minds of a group of communicating persons. In this connection, concepts must be formed in this mind (these minds), in particular, what I have called accessible theoretical variables. In the present paper, these variables, in agreement with the models above, will be physical variables like space, time, mass, momentum, charge, spin component etc., that, in most cases may be said to have an existence related to an objective reality, at least in classical theories, but in connection to measurements, to theory building and theory assessments, the variables must also be said to exist in the mind of some person and/or in the joint minds of a group of communicating persons. He/they may simply think of these variables.

An alternative option may be to just see physical variables as `stand-alone' objects connected to some established mathematical model.  Regardless of how we regard these variables, we need a measurement theory. It is argued in [5] that a quantum theory of measurement is much easier to understand if one takes as a basis the version of variables that exist in our minds.

In chapter 4 of [5] and in [6], essential elements of quantum mechanics are deduced from some concrete theorems regarding these theoretical variables. Our task here will be to try to connect such variables to relativity theory as well, and to look at some consequences of such connections.

\section{Quantum theory from theoretical variables}
\label{sec2}

This author has a background as a statistician, but from this background he has worked with the foundation of quantum mechanics for many years. The result of this work is the book [5] and several published papers in physics journals and on the arXiv. The work is now summarized in [6].

One main result from [5], as generalized in [6], is the following:
\bigskip

\textbf{Theorem 1} \textit{Consider a situation where there are two maximal accessible theoretical variables $\theta$ and $\xi$, which are real-valued or real vectors. Make the following assumptions:}

\textit{ (i) On one of these variables, $\theta$, there can be defined group actions from a transitive group $G$ with a trivial isotropy group and with left invariant measure $\rho$ on the space $\Omega_\theta$.}

\textit{ (ii) There exists a unitary multivariate representation $U(\cdot )$ of the group $G$ defined on $\theta$ such that the coherent states $U(g) |\theta_0\rangle$ are in one-to-one correspondence with $g\in G$ and hence with the values of $\theta$.}

\textit{ (iii) The two maximal accessible variables $\theta$ and $\xi$ can both be seen as functions of an inaccessible variable $\phi$: $\theta=v_\theta(\phi)$ and $\xi=v_\xi (\phi)$. There is a transformation $k$ acting on the space $\Omega_\phi$ such that $v_\xi (\phi)=v_\theta (k\phi)$.}

\textit{Then there exists a Hilbert space $\mathcal{H}$ connected to the situation, and to every accessible theoretical variable there can be associated a unique symmetric operator on $\mathcal{H}$.}
\bigskip

Of course the Hilbert space $\mathcal{H}$ here is the one associated with the representation (ii) in the theorem. The most important result is that for every accessible theoretical variable there is an associated unique operator on this Hilbert space. Explicit formulas for the operators of $\theta$ and $\xi$ are given in [5], [6] and [7]. To find the operators of other accessible theoretical variables, we can use the spectral theorem.

To understand this theorem, some definitions are necessary; for these, see the Introduction above. 

To repeat: Mathematically, to prove the theorem, we only need the following conditions: If $\lambda$ is a theoretical variable, then $\theta =f(\lambda)$ for some fixed function $f$, then $\theta$ is also a theoretical variable. And if $\lambda$ is accessible, then also $\theta$ is accessible. But in the interpretation of the theorem, I choose to connect the variables to the mind of an observer or to the joint minds of a group of communicating observers. The notion of a theoretical variable can then be seen as a generalization of the statistician's parameter notion. As such it is crucial, at least in the continuous case, to distinguish between data and variables. Future data can be modelled as (theoretical) variables plus random noise.

It is important that, in situations related to quantum theory as approached in [5], inaccessible theoretical variables also exist, like the full spin vector of a particle or the vector (theoretical position, theoretical momentum). However, these are not assumed to take sharp values. Thus Heisenberg's uncertainty relation is an important assumption behind the theorem, essentially the only physical assumption that is needed.

The assumption (iii) can be satisfied under weak conditions. When it is satisfied, we say that the variables $\theta$ and $\xi$ are \emph{related}. When no such sets of functions can be found, we say that $\theta$ and $\xi$ are essentially different.

The operators of related theoretical variables have a close relationship. To formulate this precisely, we first need a definition.
\bigskip

\textbf{Definition 1} \textit{The accessible variable $\theta$ is called permissible with respect to the group $K$ acting on $\Omega_\phi$ if the following holds: $v_\theta(\phi_1)=v_\theta(\phi_2)$ implies $v_\theta(t\phi_1)=v_\theta(t\phi_2)$ for every $t\in K$. }
\bigskip

This notion is studied thoroughly in [8]. The main conclusion is that if $\theta(\cdot)$ is permissible, then there is a group $G$ acting on the image space $\Omega_\theta$ such that $g(\theta(\phi))$ is defined as $v_\theta(k\phi)$; $k\in K$. The mapping here from $K$ to $G$ is a homomorphism. If $K$ is a transitive group on $\Omega$, then $G$ is transitive on $\Omega_\theta$. (See Lemma 4.3 in [5].) 
\bigskip

\textbf{Theorem 2}
\textit{Assume that the function $\theta(\cdot)$ and $\xi(\cdot)$ are permissible with respect to a group $K$ acting on $\Omega$. Assume that $K$ is transitive and has a trivial isotropy group. Let $T(\cdot)$ be an irreducible unitary representation of $K$ such that the coherent states $T(t)|\psi_0\rangle$ are in one-to-one correspondence with $t$. For any transformation $t\in K$ and any such unitary representation $T$ of $K$, the operator $T(t)^\dagger A^\theta T(t)$ is the operator corresponding to $\theta'$, given by $\theta'(\phi)=\theta(t\phi)$.}

\textit{In addition, there is a unitary operator $W$ on the Hilbert space such that the operators associated with $\theta$ and $\xi$ are connected by $A^\xi = W^\dagger A^\theta W$.}
\bigskip

Theorem 2 is proved in the Appendix of [6].

In Chapter 4 of [5], and also in [6], it is proved that, in the discrete case, essential elements of ordinary quantum mechanics follow from variants of Theorem 1 and Theorem 2 above. In this case, it is also shown explicitly how $G$, $k$, and $K$ can be chosen. Thus the theory simplifies considerably in the finite-dimensional case: No symmetry assumption is needed. By only assuming that $\theta$ and $\xi$ both take $n$ values and are maximal, the whole Hilbert space formalism results.

In general, the assumption that there can be defined a transitive group $G$ acting upon $\theta$ is crucial. It can easily be satisfied when the range of $\theta$ is finite or is the whole line ${\mathbb{R}}^1$, but is also relevant when $\theta$ is a vector. As an example, assume that $\theta$ takes all values in some Euclidean space ${\mathbb{R}}^p$. Then all the necessary assumptions are satisfied by the translation group: $\theta\mapsto\theta+\alpha$, where $\alpha$ is some arbitrary vector in ${\mathbb{R}}^p$. In the finite case, the group $G$ can be taken to be the cyclic group on $\Omega_\theta$. When $\theta$ and $\theta'$ take two values, say $-1$ and $+1$, they can be taken to be spin components, and the group $K$ can be defined as the group of rotations in the plane determined by the two components.

Thus Theorem 1 and Theorem 2 can be used in a new foundation of quantum theory, and this foundation is in no way limited to finite-valued or scalar variables.

However, in the discrete case, more can be proved [5, 6]: The set of eigenvalues of the operator $A^\theta$ equals the set of possible values of $\theta$. The accessible variable $\theta$ is maximal if and only if each eigenvalue is single that is, each eigenspace is one-dimensional. This gives a nice interpretation of the eigenvectors of operators with a physical interpretation. In the present article I will limit the concept of state vectors to vectors that can be given such an interpretation. It is shown in [6, 7] that certain entangled states may also be interpreted in this way.

In [6], all the mathematical proofs are now collected. 

In the finite-dimensional case, the eigenspaces of the operators connected to variables $\lambda$ are in one-to-one correspondence with questions: `What will be the value of $\lambda$ if I measure it?', together with a sharp answer `$\lambda=u$'. If and only if the accessible variable $\lambda$ is maximal, the eigenspaces are one-dimensional. This gives a concrete, very simple interpretation of many unit vectors in the Hilbert space. The difficult problem of determining when all relevant unit vectors in some concrete situation can have such an interpretation is briefly taken up in the technical paper [33].

Having established this important foundation, the main other foundational result to prove is the Born formula. In [5] and [6], this formula is proved under the following three assumptions: 1) The likelihood principle from statistics holds (this principle is motivated in Chapter 2 of [5]). 2) The actor performing the relevant experiment or measurement has ideals that can be modelled by a perfectly rational abstract being. 3) The state in the mind of this actor describing the physical system before the measurement or experiment is coupled to a maximal accessible variable. It can be shown [6] that the Born formula can be given a form where the last assumption can be dispensed with, but then we have to assume that the relevant theoretical variable $\theta$ is dominated by a maximal accessible variable $\eta$ such that the conditional distribution of $\eta$, given $\theta$ is uniform.

In [5] the derivation of the Schr\"{o}dinger equation is also taken up.

One can also, from this basis discuss several so-called paradoxes of quantum mechanics. In particular, in connection to the Schr\"{o}dinger cat paradox, it is argued for a version of quantum theory where the state vector concept is limited to eigenvectors of physically meaningful operators. In this version it is possible to link all pure states to question-and-answer pairs as above.  In the present article I will limit my discussion of quantum theory to situations where the above link can be assumed. I will also assume that the questions above in certain cases can be answered by `I don't know', and I will assume an epistemic interpretation of quantum mechanics.

In this version linear combinations of states are limited to combinations of the form
\begin{equation}
|b\rangle =\sum_j | |a,j\rangle\langle a, j ||b\rangle =\sum_j \langle a, j|b\rangle |a,j\rangle ,
\label{linear}
\end{equation}
for a complete orthonormal set of states $\{|a,j\rangle \}$.
\bigskip

\textbf{Example 1.} \textit{Schr\"{o}dinger's cat.} The discussion of this example concerns the state of the cat just
before the sealed box is opened. Is it half dead and half alive?

To an observer outside the box the answer is simply: ``I do not know''. All accessible theoretical variables
connected to this observer are free of any information about the status of life of the cat. But on
the other hand – an imagined observer inside the box, wearing a gas mask, will of course know the
answer. The interpretation of quantum mechanics is epistemic, not ontological, and it is connected to the
observer. Both observers agree on the death status of the cat once the box is opened.
\bigskip

\textbf{Example 2.} \textit{Wigner’s friend.} Was the state of the system only determined when Wigner learned the
result of the experiment, or was it determined at some previous point?

My answer to this is that at each point in time a quantum state is connected to Wigner’s
friend as an observer and another to Wigner, depending on the knowledge that
they have at that time. The superposition given by formal quantum mechanics corresponds to a `do not
know' epistemic state. The states of the two observers agree once Wigner learns the result of the
experiment.
\bigskip

\textbf{Example 3.} \textit{The two-slit experiment.} This is an experiment where all real and imagined observers
communicate at each point in time, so there is always an objective state. 

Look first at the situation when we do not know which slit the particle goes through. This is 
a `do not know' situation. Any statement to the effect that the particles somehow pass through both
slits is meaningless. The interference pattern can be explained by the fact that the particles are (nearly)
in an eigenstate in the component of momentum in the direction perpendicular to the slits in the plane
of the slits. And by de Broglie's theory, momentum is connected to a wave. On the other hand, if an observer finds out which slit the particles go through, the state changes into an eigenstate for position in that direction. In either case, the state is an epistemic state for each of the communicating observers, which might indicate that it in some sense can be seen as an ontological state. But this may be seen as a state of the screen and/or the device to observe the particle, not as an ontological state of the particle itself.
\bigskip

For further consequences of this theory, see [5] and [6].

\section{Causality, inference, and reality}
\label{sec3}

The book [5] concentrates on epistemic processes, processes to obtain knowledge through experiments or measurements. (Of course, there are also other ways to obtain knowledge; this is largely ignored in [5].) A very important problem that remains to be discussed is to what extent the results of such epistemic processes can be associated with some sort of reality, a `real' world. The only statement about this given in [5] is the following: `If all real and imagined observators can be said to agree on the result of some experiment or measurement, then this is a strong argument to the effect that this result can be coupled to some reality. This conclusion is strengthened if the experiment is done in a `proper' scientific way.'

Recently, a deeper discussion of the reality question was attempted by Schmid et al.  [12]. Two weaknesses of that paper, however, are first that statistical inference is limited to Bayesian inference, and next that the paper in some sense mixes the concept of ontology with something related to cause-and-effect relations. Another answer to the question of whether classical ontology can be made compatible with quantum mechanics is given by Evans [13], a paper where [12] is criticized as well. My own views on this question are now given in [14].

\section{Theoretical variables related to relativity theory}
\label{sec4}

It is sometimes said that one of the obstacles to combining quantum theory and general relativity theory is that in quantum field theory [15], say, time and space are independent variables, while in general relativity theory, time and space are the basic constituencies. I want to tune down this difference here: To me, here \emph{time and space are physical variables, but also theoretical variables associated with the mind of some actor or to the joint minds of a group of actors.} In a given situation, some actors may focus on the theories where time and space are independent variables (relativity theory), while other actors may focus on theories like quantum field theories. Most people do not focus on any of these theories at all, but researchers trying to think deeply do. In the following I will not rely on any of the deep theoretical variables that recent researchers have invented in attempts to understand the general situation. In particular, I will not mention strings, loops, or multiple universes. I will only take as my points of departure simple variables, in particular space and time, momentum and energy. It is interesting, however, to ask whether my approach towards quantum mechanics can be generalized to modern quantum field theories in the way these theories are developed as a background for the standard model in physics.

\section{Field theories and gauge groups}
\label{sec5}

I will start with classical field theories, where by `classical' I also include special and general relativity theory. The field theories will be seen as models in physics, and as such, they also exist in the joint minds of a communicating group of physicists. In a concrete setting, important variables are space and time. A concrete event can always be thought of as taking place at a specific time-space point $\tau=(t,x,y,z)$, where $t$ is the time as measured by some actor, and $(x,y,z)$ are the space variables as measured by the same actor. In general, $\tau$ is a theoretical variable, and it varies in, say $\Omega_\tau$. 

A field is then defined as a function from $\Omega_\tau$ to another mathematical space $\Omega_\psi$:
\begin{equation}
\tau\mapsto\psi(\tau).
\label{a}
\end{equation}

In agreement with my previous theory, I assume that some fields are accessible. Physical examples are electrical and magnetic fields. As my first basic model, I assume the existence of a large inaccessible field $\phi=\phi(\tau)$ such that all accessible ones are functions of this field, say $\theta(\tau)=f_\theta (\phi(\tau))$. I assume just that $\phi$ takes values in some mathematical space $\Omega_\phi$. But I also assume that a group $K$ is defined acting on $\Omega_\phi$. If $\theta(\cdot)$ defined by $\tau\mapsto\theta(\tau)$ is accessible, then $G$ is assumed to be a transitive group acting on $\Omega_\theta$. It may or may not be that the function $f_\theta$ is permissible with respect to $K$. If it is permissible, then $G$ may be defined by $gf_\theta(\phi(\tau))=f_\theta(k\phi(\tau))$ for $k\in K$.

A local variant of this will appear if the group elements $g\in G$ and $k\in K$ depend on the time-space point $\tau$, and perhaps also $f_\theta$ depends on $\tau$. In the global case also $\phi(\cdot)$ is independent of $\tau$.

The field $\phi$ defined above must be considered just as a mathematical variable; it can have no physical interpretation. The whole approach corresponds to Section 3 in [6]. In Section 4 there, I concentrate on physical variables, and they are all accessible. Then I build on some simple category theory, and I replace $\Omega_\phi$ with a basic underlying object $\Omega$, interpreted in [6] as a space of notions connected to an observer or joint notions for a set of communicating observers. The group $K$ is now a group of automorphisms on $\Omega$, assumed to exist. The basic assumption is the following: \emph{For each accessible physical field $\theta(\tau)$ there exists a set of morphisms $v_\theta^\tau$ from $\Omega$ onto $\Omega_\theta$ and at least one $u=u^\tau \in \Omega$ such that $\theta(\tau)=v_\theta^\tau( u^\tau)$.}

In the first case, a quantum version of the field theory may tentatively be defined by appealing to Theorem 1 and Theorem 2 above. Versions of these theorems for the second model are also given in [6]. The basic assumption behind these versions of Theorem 1 is that we have two related maximal accessible fields $\theta(\tau)$ and $\xi(\tau)$, and that the group $G$ acting upon $\Omega_\theta$ has certain properties. Specifically it should be transitive and have a trivial isotropy group, and it should have an irreducible representation $U(\cdot)$ such that the coherent states $U(g)|\theta_0\rangle$ for some fixed state vector $|\theta_0\rangle$ are in one-to-one correspondence with $g$.

If this is the case, quantum operators $A^{\theta(\tau)}$and $A^{\xi(\tau)}$ can be defined for each $\tau$. In the global case, these operators will be independent of $\tau$; in the local case, they will depend on $\tau$.

Here, I will not go into concrete applications of this other than some associated with relativity theory, but I will define in general what I mean by a gauge group.
\bigskip

\textbf{Definition 2}
\textit{The gauge group is a subgroup $H$ of the group $K$, and it is defined with respect to all (maximal) accessible fields and variables $\theta(\cdot), \xi(\cdot), \lambda(\cdot)...$. Specifically, it is defined as the maximal group such that for each $\tau$ all $\theta(\tau), \xi(\tau), \lambda(\tau)...$ are constant; in the first model: $f_\theta(h\phi)=f_\theta(\phi)$ and so on.}

\textit{As before, we can have a local variant where the elements $h$ depend on the time-space variable $\tau$.}
\bigskip

It is enough to verify the criterion of constancy for the maximal variables. And if the maximal variables are related, it is enough to verify this criterion for one variable.

Note that a change of gauge $h$ will not affect any accessible variables, so the physics will be the same. Gauge theories are central in modern physics, in particular in connection to quantum field theory; I will not go further into any of these themes here, but refer to [15]. However, I am interested in a possible gauge theory associated with special and general relativity; this will be very briefly discussed later, but it is convenient to introduce a Lagrangian density and a Lagrangian already here.

 In the first model, assume that derivatives with respect to the four-vector $\tau$ can be defined in the space $\Omega_\phi$, and let the components be $\partial_\mu\phi$ for $\mu=1,...4$. Denote the space in which this four-vector varies as $\Omega_\pi$, and define $\Omega_\psi =\Omega_\phi \otimes \Omega_\pi$. The Lagrangian density is then defined as some function on $\Omega_\psi$, and the Lagrangian as the integral of this function over four-space, which can be seen as a function on the field $\psi(\cdot)$. I will not consider here how to generalize to the second model.
 
 In order to include the Lagrangian, we now extend Definition 2 to the field $\psi(\cdot)$, and assume that the group $K$ can be defined to act on the whole space $\Omega_\psi$. The accessible variables $\theta(\tau), \xi(\tau), \lambda(\tau)...$ may also be defined on $\Omega_\psi$ in general.

\section{Information and entropy}
\label{sec6}

Since Shannon [16[, information has been coded in bits. However, the term information also has wide connotations: one person can have information about other persons or about phenomena in the real world. In his mind, this is coded in terms of theoretical variables. In this article, I will consider a situation where an actor or a group of communicating actors have focused on one particular maximal accessible variable $\theta$. The information connected to this situation can be formulated in terms of the bits associated with the different values of $\theta$. Note again that this information depends upon the particular actor/ group of actors. In a concrete relativistic setting, $\theta=\tau$ may for instance be the spacetime values connected to some physical system, and we may also be interested in the complementary variable $\xi$, the energy-momentum vector connected to the physical system. The information associated with these complementary variables will in general be different. We consider first the case where $\theta$ and $\xi$ take a finite or countable set of values.

The Shannon information associated with a variable assumes a probability distribution over this variable. We will assume that our knowledge of the maximal accessible variable $\theta$ is given by a pure state which alternatively can be described by the density operator
\begin{equation}
\rho =\sum_i p_i P_i ,
\label{i}
\end{equation}
where $P_i =|\psi_i\rangle\langle \psi_i |$ are orthogonal one-dimensional projection operators, and $p_i$ are probabilities. Then the Shannon information is given by
\begin{equation}
H^\theta =-\sum_i p_i \mathrm{log}( p_i ),
\label{ii}
\end{equation}
where the logarithm is with respect to the basis 2.

Assume now a complementary, maximal accessible variable $\xi$, which through Theorem 1 is associated with an operator $A^\xi=\sum_j a_j Q_j$, where the $Q_j =|\phi_j\rangle\langle\phi_j|$ constitute another orthogonal set of one-dimensional projection operators. Then through Born's rule the probabilities of the different values are
\begin{equation}
q_j =\langle \phi_j |\rho |\phi_j\rangle ,
\label{iii}
\end{equation}
and the associated Shannon information is
\begin{equation}
H^\xi =-\sum_j q_j \mathrm{log}(q_j).
\label{iv}
\end{equation}

Analogous formulas hold for the continuous case. For a random variable with probability density $f(x)$, the Shannon information is
\begin{equation}
H= -\int_x f(x) \mathrm{log} (f(x))dx.
\label{v}
\end{equation}

From a physical point of view, it is very important that Shannon's information is proportional to the thermodynamic concept of entropy. The proportionality constant is Boltzmann's constant $k_B$ (when natural logarithms are used). This connection was first made by Ludwig Boltzmann, and expressed by his equation for entropy
\begin{equation}
S=k_B \mathrm{ln}(W),
\label{vi}
\end{equation}
where $W$ is the number of microstates that can give a given macrostate. It is assumed that each microstate is equally likely, so that the probability of a given microstate is $p_i = 1/W$.

According to Jaynes [17], thermodynamic entropy, as explained by statistical mechanics, should be seen as an application of Shannon's information theory: The thermodynamic entropy is defined as being proportional to the amount of further information needed to define the detailed microscopic state of the system. Adding heat to a system increases the thermodynamic entropy because it increases the number of possible microstates. Maxwell's demon can hypothetically reduce the thermodynamic entropy of a system by using information about the states of individual molecules, but as shown by Landauer [18] and later coworkers, to function, the demon himself must increase the thermodynamic entropy in the process by at least the amount of Shannon information the demon uses in the process. Landauer's principle imposes a lower bound on the amount of heat a computer must generate to process a given amount of information.

Information systems have been studied by several authors. As an example, I will mention the article [19] by Liang et al., who obtain relationships between information entropy and certain knowledge measures.

It is a basic principle of physics that the entropy of a closed system can never decrease.
\bigskip

\underline{Example}

Imagine two scientists $A$ and $B$, both busy with writing articles on thermodynamics, information and entropy. Both want to illustrate their theories with an example based on a deck of 52 cards. They have certain ideas on the random process of shuffling the deck, and they want to describe these ideas in some detail. At the same time, they want to illustrate the physical law that entropy increases. So both say that the shuffling process starts in a state with low entropy: The deck is ordered. However, there is a difference between $A$ and $B$. To $A$, an ordered deck starts with the ace of spades, then the two of spades, all the other spades, then the hearts, then the diamonds, and finally all the clubs. To $B$, an ordered deck starts with all the aces, then all the two's, the three's and so on.

My point is that the description of the state of the deck after one shuffling, will not be the same for $A$ and for $B$. In a way, the state concept depends upon the actor. In the same way I will say that the entropy in physics in principle must be based on a concept of states that may depend on an actor, an observer.
\bigskip

In 1927, John von Neuman proposed a formula for the entropy connected to a quantum mechanical mixed state $\rho$:
\begin{equation}
S= -k_B \mathrm{trace}(\rho \mathrm{ln}(\rho)).
\label{vii}
\end{equation}
Apart from the constant $k_B$ and the base of the logarithm, this equals (\ref{ii}) when the state is given by (\ref{i}). It is connected to the distribution of the maximal accessible variable $\theta$ that the actor in question has knowledge of. Any other, complementary variable $\xi$ will be connected to another entropy (compare (\ref{iv})), but it is given by similar formula.

Algorithmic randomness is defined by the size in binary digits of the shortest message that can reproduce the microstate of a system uniquely in some given setting. This definition was used by Zurek [20] to discuss algorithmic randomness to measure disorder without any recourse to probabilities. Gibbs and Boltzmann's entropy, as well as Shannon's information theoretic entropy then provide estimates of the expected value of the algorithmic randomness.

In [21], there is a thorough comparison between on the one hand Kolmogorov's fundamental concept of complexity, which is the length in bits of the shortest computer program that prints a given sequence of symbols and then halts, and Shannon's concept of information on the other hand. Although their primary aim is quite different, and they are functions defined in different spaces, there is a close relationship between the two concepts. It is also pointed out that there is a relationship to the statistical notion of sufficient statistics.

Shannon's information has two interpretations: one axiomatic connected to $H$ as a function of probabilities, and one coding interpretation. The latter derives from entropy as the minimum average length in bits needed to encode outcomes in some sample space. There is also a connection between Shannon information and Kolmogorov complexity: Expected Kolmogorov complexity equals Shannon entropy. Both concepts lead to a notion of mutual information $I$ between two variables $ \theta$ and $\xi$:
\begin{equation}
I(\theta,\xi)=H(\theta)-H(\xi |\theta).
\label{vii}
\end{equation}

In a statistical setting one can talk about the mutual information between data $x$ and parameter $\theta$, related to the probabilistic model for data, given a parameter and a possible prior for this parameter. A function of data $S(x)$ is \emph{sufficient} relative to the model iff
\begin{equation}
I(\theta, x)=I(\theta, S(x))
\label{viii}
\end{equation}
for all prior distributions of $\theta$. This is equivalent to
\begin{equation}
H[x|\theta)=H(S(x)|\theta)
\label{ix}
\end{equation}
for all $\theta$.

\section{Special relativity}
\label{sec7}

Both special and general relativity theories discuss how variables change when the observers change. For space and time this is essential: Special relativity theory is concerned with observers that move with a uniform speed relative to each other; in general relativity theory relative acceleration is allowed. 

However, none of these theories take up the problems associated with the fact that concepts can be related to the minds of people. Here I want to discuss some aspects of this. I will first concentrate on special relativity theory.

Take as a point of departure an observer $A$ with space coordinates $(0,0,0)$, and let the time $t$ run. Relative to this observer, a given physical system, say, a particle, may be characterized by special values of the four-vector $\theta=(t,x,y,z)$, which for instance may give the location of the particle at time t for this observer. This may be accessible at some fixed time $t$, but is inaccessible as a process. Alternatively, one may look upon the `particle' as a wave, specify its frequency $f$ and its wavevector $k$, hence its energy $E=hf$ and its momentum $p=hk$, that is, values of the four-momentum $\xi =(E,p_x ,p_y ,p_z )$. Both $\theta$ and $\xi$ are maximal accessible variables and can be seen as physical variables, but may also be associated with the mind of the observer $A$. We can show that by Theorem 1, these two variables imply a Hilbert space $\mathcal{H}$, and on this Hilbert space, $\theta$ has an operator $A^\theta$, and $\xi$ has an operator $A^\xi$. Both these operators change when the observer changes.

Crucially for the proof of Theorem 1 are the definition of the group $G$ on the $\theta$-space, the definition of the inaccessible variable $\phi$, and the construction of a suitable transformation $k$ in the $\phi$-space. For $G$, we may take the group of 4-dimensional translations. We can just take $\phi = (\theta, \xi)$ and let $k$ be a suitable element of the Weyl-Heisenberg group acting on $\phi$.

It is also crucial here that $\theta$ and $\xi$ can be looked upon as theoretical variables, not necessarily data. Earlier, all theoretical variables were denoted by Greek letters; it is hoped that the Latin letters above do not lead to any misunderstanding. It is assumed that the \emph{measurement} of any function of $\theta$, say the $x$-coordinate, can be modelled by the theoretical variable $x$ plus some random noise.

Note that the Poincar\'{e} group $P$ also can be seen as acting on the four-momentum $\xi$.

Let $B$ be an observer that moves relative to $A$ with a constant speed $v<c$. Then both $\theta$ and $\xi$ change according to actions of the Poincar\'{e} group $P$. This group is transitive on the relevant spaces. Its group elements $p$ can be seen as a combination of a translation $g$ and a member $l$ of the Lorentz group. The Lorentz group in turn consists of rotations and Lorentz boosts, coordinate frames moving with constant velocity along the positive $x$-axis.

Assume that the given event is in the future light cone both for $A$ and for $B$ at time 0. The clocks are calibrated such that $t=0$ for $A$ coincides with $t'=0$ for $B$.

Unitary representations of the translation group $G$ are discussed in textbooks. I will not go into details here. It suffices to say that a representation $U(g)$ can be found that the coherent states $U(g)|\psi\rangle$ are in one-to-one correspondence with the group elements $g$. Thus from Theorem 1 operators $A^\theta$ and $A^\xi$ acting on a suitable Hilbert space may be constructed. 

If $p$ is the element of the Poincar\'{e} group transforming $\theta$ for observer $A$ into the corresponding coordinate $\theta'$ for $B$, then $A^{\theta'}=V(p)^\dagger A^\theta V(p)$, where $V$ is a unitary irreducible representation of the Poincar\'{e} group. Such representations were discussed by Wigner in 1939 [22]. 

To study the change of the operator $A^\xi$ when $A$ is replaced by $B$, we first need some group element $k$ in a larger group $K$ acting on the vector $\phi=(\theta,\xi)$ such that $(\xi,\theta) = k(\theta, \xi )$. This can be achieved by considering a variant of the Weyl-Heisenberg group connected to the observer $A$. Let $p$ in the Poincar\'{e} group also be seen as a member of the large group $K$ by $p(\theta,\xi)=(p\theta,\xi)$. Then $(\theta',\xi')=h(\theta',\xi)$ is found from $h=kpk$ since $k^2$ is the identity. By Theorem 2, we then get $A^{\xi'}=T(h)^\dagger A^\xi T(h)$ for some unitary irreducible representation $T$ of the large group $K$. It is left to prove that the relevant functions are permissible with respect to the group $K$, but this I will leave as an open mathematical problem.

Operators associated with groups can be constructed in many ways. One well-known is as generators of Lie algebras connected Lie groups. This approach is taken in [23] for the Poincar\'{e} group and several related groups. An interesting feature is that all the groups are derived there through symmetries of the commutation relations associated with Heisenberg's uncertainty relations.

But go back to the two observers and their theoretical variables in the way this was introduced above. We have two possibilities: The relationship between the observers may be timelike or it may be spacelike. In the first case, assume that $B$ is in the future light cone for $A$. Both observe the event given by $A$ as happening at time $t$ and space coordinates $(x,y,z)$. Both have in principle two possibilities: They can measure $\theta=(t,x,y,z)$, respectively $\theta'$, or they can measure the complementary variables $\xi =(E,p_x,p_y,p_z)$, respectively $\xi'$.

Look first at the timelike case. Assume an ideal situation such that $A$ immediately after his measurement is able to send his result to $B$ with a light signal travelling with speed $c$. Then $B$ knows the value of either $\theta$ or $\xi$, and by using his knowledge of the Poincar\'{e} transformation, he can find the corresponding $\theta'$, respectively $\xi'$. By Heisenberg's uncertainty relation he is not allowed to know both these variables exactly. But that must mean that he at the same time is not able to choose to measure the other variable. Hence we seem to conclude, by using both relativity theory and quantum mechanics in our reasoning, that in this case, $B$ is limited in his choice of measurement. However, this contradicts the axiom of free choice. Hence one must modify the reasoning behind this paradox. The simplest modification is to assume that $A$ always must have a shorter or longer delay in time from the moment he receives the result of his measurement to the moment when he is able to send his results away.

In the other case, when $A$ and $B$ have a spacelike separation, they are not able to communicate, and we are not able to use the argument above. Hence in this case it is clear that $B$ can choose his measurement freely.

Next, have a brief look at a kind of gauge group connected to relativity theory. Let us take the point of view that the combined set of laws of physics constituts our accessible `variables'. Since the Lagrangian of some systems determines the dynamics, this must mean in particular, that the Lagrangian is accessible. As noted in Section 5, the gauge group is the group $H$, where all the accessible variables are constant.

The group $K$ acting on $\phi = (\theta,\xi)$ can be taken to be a four-dimensional version of the Weyl-Heisenberg group. This group is transitive and has a trivial isotropy group. As a starting point, I take the Lagrangian $\lambda$ to be a function of $\phi$, assumed to be permissible with respect to $K$. Then this induces a group $L$ acting on $\lambda$. 
The property of permissibility implies the following: The inverse image of the function $\lambda(\cdot)$ induces a subgroup $K^L$ of $K$, and this, following the definition in Section 5, will be the relevant gauge group. In simpler terms, we can write $K^L = K/L$.

 However, $\theta$ and $\xi$ are also accessible variables, and the groups associated with these are two four-dimensional translation groups $T$ and $S$. So this should imply that the resulting gauge group can be taken to be $H=K/(L\otimes T\otimes S)$. I assume here that both $\theta(\cdot)$ and $\xi(\cdot)$ are permissible with respect to $K$.

More realistic gauge theories assume a Lagrangian which also depends on space and time derivatives of the field $\phi$. Then one has to introduce the larger space $\Omega_\psi$ defined in Section 5, and let $K$ be a group acting upon this space. If, again, the Lagrangian $\lambda(\cdot)$ can be seen as a permissible function with respect to $K$, the gauge group can be defined as before. This gives a global gauge theory.

In the arguments above I have referred to special relativity theory. However, parts of the arguments can be extended to a more general case.

\section{General relativity; a summary}
\label{sec8}

The core of general relativity theory is the equivalence principle: Seen locally, an observer in a closed box is not able to distinguish between the effect of gravity and the effect of acceleration. One consequence of this is that, locally, one can always choose at least one coordinate system such that, with respect to this coordinate system, the laws of special relativity hold. 

But this can, in principle, be used to construct a local gauge theory for general relativity, also. Let again $K$ be a group acting upon the space $\Omega_\psi$, and let $K^L =K/L$ now be the subgroup where the Lagrangian density $\lambda$ is constant. Fix a time-and-space vector $\tau=\theta$, and let $S$ be the translation group in four-momentum $\xi$. Assume that both $\lambda(\cdot)$ and $\xi(\cdot)$ are permissible with respect to $K$. Then, a local gauge group may be taken as $H=K/L\otimes S$.

A technical problem here might be to construct a version of general relativity based upon waves as input instead of spacetime. (Energy and momentum are determined from the wave.) To proceed with this problem, let us assume a theory based upon waves with frequency $f$ and wave vector $k$, equivalently on the energy $E=hf$ and the momentum $p=hk$, hence based on $\xi =(E,p_x,p_y,p_z)$. Assume that the Lagrangian density can be found as a function of $\xi$ and the partial derivatives with respect to $\xi$. This gives a local gauge group for general relativity as above.

The gauge theory of general relativity is a continuum field theory. I will not go into details here but refer to the literature.

Central to general relativity is the metric tensor $\bm{g}=\{g_{\alpha\beta}\}$. In the special local coordinate system where the laws of special relativity hold, this can be taken as
\begin{equation}
\bm{g}=\bm{\eta}=\left(
\begin{array}{cccc}
 -1&0&0&0\\0&1&0&0\\0&0&1&0\\0&0&0&1
 \end{array}\right).
\label{gr1}
\end{equation}

In a general coordinate system, say $\{\theta\}$,  the metric matrix $\bm{g}=(g_{\alpha\beta})$ can be any symmetric matrix with trace $2$ which can be changed by a local coordinate transformation to $\bm{\eta}$.

For change of coordinates from $\{\theta\}$ to $\{\theta'\}$ in this space introduce the Jacobian tensor $\bm{\Lambda}=\{\Lambda^{\alpha}_{\ \beta'}\} = \partial \theta/\partial\theta'$. Then
\begin{equation}
\bm{g'}=\bm{\Lambda}^T\bm{g}\bm{\Lambda},
\label{gr2}
\end{equation}
or, by using the common summation convention:
\begin{equation}
g'_{\alpha'\beta'} =\Lambda_{\alpha'}^{\ \mu}g_{\mu\nu}\Lambda^{\nu}_{\ \beta'}.
\label{gr3}
\end{equation}

For a moving particle with coordinates $\theta$, it is convenient to introduce a proper time
\begin{equation}
d\tau^2 = -g_{\alpha\beta}\theta^\alpha \theta^\beta,
\label{gr32}
\end{equation}
and also the four-velocity $U_\alpha = d\theta_\alpha/d\tau$ and the momentum $p_\alpha =m U_\alpha$, where $m$ is the  mass. 

For any vector $\bm{V}$ in any coordinate system we have
\begin{equation}
V_\alpha =g_{\alpha\beta}V^\beta,\ \ \ V^\alpha=g^{\alpha\beta}V_{\beta}
\label{gr31}
\end{equation}
and similarly for tensors.

Furthermore, by change of coordinates
\begin{equation}
V'_{\alpha'}=\Lambda_{\alpha'}^{\ \beta}V_\beta .
\label{gr30}
\end{equation}

The partial derivative of a vector is denoted by a comma, and is defined by the partial derivative of each component:
\begin{equation}
V_{\alpha,\mu}=\partial V_\alpha/\partial\mu .
\label{gr31}
\end{equation}

In a similar way one can define the derivative of any scalar or tensor. The derivative of the determinant $g$ of the matrix $(g_{\alpha\beta})$ is
\begin{equation}
g_{,\mu}=gg^{\alpha\beta}g_{\beta\alpha,\mu}.
\label{gr6}
\end{equation}

Basis vectors are denoted by $e_\alpha$, and derivatives of these by $e_{\alpha,\nu}=\partial e_\alpha /\partial \theta^\nu$. This leads to the important Christoffel symbol defined by
\begin{equation}
e_{\alpha,\beta} =\Gamma^{\mu}_{\ \alpha\beta}e_{\mu}.
\label{gr4}
\end{equation}
One can show [24] that one always have $\Gamma^{\mu}_{\ \alpha\beta}=\Gamma^{\mu}_{\ \beta\alpha}$, and
\begin{equation}
\Gamma^{\mu}_{\ \alpha\beta}=\frac{1}{2}g^{\mu\nu}(g_{\nu\alpha,\beta}+g_{\nu\beta ,\alpha}-g_{\alpha\beta,\nu}).
\label{gr5}
\end{equation}

For any vector or tensor one can define covariant differentiation taking into account the derivatives of the unit vector. For instance
\begin{equation}
T^{\alpha\beta}_{\ \ ;\gamma}=T^{\alpha\beta}_{\ \ ,\gamma}+\Gamma^{\alpha}_{\ \mu\gamma}T^{\mu\beta}+\Gamma^{\beta}_{\ \mu\gamma}T^{\alpha\mu},
\label{gr7}
\end{equation}

while for a vector
\begin{equation}
V^{\alpha}_{\ ;\mu}=V^{\alpha}_{\ ,\mu} +\Gamma^{\alpha}_{\ \mu\nu}V^{\nu}.
\label{gr70}
\end{equation}

The process of going from commas to semicolons is important in deriving equations of general relativity. For instance, if we know for a vector $\bm{V}$ that $V^{\mu}_{\ ,\mu}=0$ holds in the special coordinate system determined locally by the metric tensor $\eta_{\alpha\beta}$, then this is equivalent in this system to $V^{\mu}_{\ ;\mu}=0$, which can be generalized to any coordinate system.

An important special tensor is the Riemann curvature tensor $\bm{R}$, which describes how a vector changes under parallel transport around a loop. It can be defined [24] as
\begin{equation}
R^{\alpha}_{\ \beta\mu\nu}=-\Gamma^{\alpha}_{\ \beta\mu,\nu}+\Gamma^{\alpha}_{\ \beta\nu,\mu}+ \Gamma^{\alpha}_{\ \sigma\mu}\Gamma^{\sigma}_{\ \beta\nu}-\Gamma^{\alpha}_{\ \sigma\nu}\Gamma^{\sigma}_{\ \beta\mu}.
\label{gr8}
\end{equation} 
Alternatively it can be defined in terms of second derivatives of the metric matrix $\bm{g}$. The tensor $\bm{R}$ is zero for a flat manifold.

Contraction of indices in $\bm{R}$ can be defined by using the summation convention. The Ricci tensor and the Ricci scalar are defined by
\begin{equation}
R_{\alpha\beta}= R^{\mu}_{\ \alpha\mu\beta}\ \ \ R=g^{\mu\nu}R_{\mu\nu}.
\label {gr9}
\end{equation}

The Riemann curvature tensor satisfies some simple identities and also the Bianchi identities:
\begin{equation}
R_{\alpha\beta\mu\nu;\lambda}+R_{\alpha\beta\lambda\mu;\nu}+R_{\alpha\beta\nu\lambda;\mu}=0.
\label{gr10}
\end{equation}

The Einstein tensor is defined by
\begin{equation}
G^{\alpha\beta}=R^{\alpha\beta}-\frac{1}{2}g^{\alpha\beta}R,
\label{gr11}
\end{equation}
and the Einstein field equations (with a vanishing cosmological constant) are now simply
\begin{equation}
G^{\alpha\beta}=8\pi T^{\alpha\beta},
\label{gr12}
\end{equation}
where $\bm{T}=(T^{\alpha\beta})$ is the so-called stress-energy tensor. 

In the frame with metric $\bm{\eta}$, $T^{\alpha\beta}$ is defined [24] as the flux of component $\alpha$ of the four-momentum $\xi =(E, p_x, p_y, p_z)$ across a surface of constant component $\beta$ of $\theta=(t,x,y,z)$. In particular, $T^{00}$ can be interpreted as the energy density. Note that this definition assumes that both $\theta$ and $\xi$ are accurately known, something that is in contradiction to quantum theory. This is a basic contradiction that must be solved in order to have a joint theory including both relativity theory and quantum theory.

Sticking with relativity theory, the definition can be extended to all coordinate systems by using a generalization of equations (\ref{gr3}) and (\ref{gr30}):
\begin{equation}
T'^{\alpha'\beta'} = \Lambda^{\alpha'}_{\ \alpha}T^{\alpha\beta}\Lambda_{\beta}^{\ \beta'},
\label{gr13}
\end{equation}
where again $\bm{\Lambda}=\partial\theta/\partial\theta'$.

From the Bianchi identities one can show
\begin{equation}
T^{\alpha\beta}_{\ \ ;\beta}= G^{\alpha\beta}_{\ \ ;\beta}=0,
\label{gr13}
\end{equation}
which is the equation of local conservation of energy and momentum.

\section{General relativity; two different observers}
\label{sec9}

Both $\theta$ and $\xi$ are four-vectors, and by a change of coordinate system, their components change according to the equation (\ref{gr30}). A particular case of this is the change of observer. One observer may use the coordinates $\theta$, the other observer the coordinates $\theta'$. The crucial tensor is then given by $\bm{\Lambda}=\{\Lambda^{\alpha}_{\ \beta'}\} = \partial \theta/\partial\theta'$.

In addition, each observer must make a choice of what variable to focus on in his experiments. By Heisenberg's uncertainty relation he cannot choose both $\theta$ and $\xi$, but must concentrate on one of them. This choice is made independently for each observer. Let observer Alice have the choice between $\theta$ and $\xi$, while observer Bob has the choice between $\theta'$ and $\xi'$. 

Let a new observer Charlie observe both Alice and Bob. We can always arrange it in such a way that Alice and Bob are in the past `light cone' of Charlie. So Charlie has all the data of all experiments made by Alice and Bob. He can try to make up a joint model describing all these experiments.

According to the analysis made in [9], the actor Charlie will be limited in his attempts to model the situation. In agreement with the simple quantum model of [6], assume that all accessible variables are functions of some underlying inaccessible variable $\phi$. An accessible variable $\eta$ is called maximal if it cannot be extended to a wider accessible variable. Two maximal accessible variables $\eta$ and $\zeta$ are said to be related if $\eta = f(\phi)$ and $\zeta = f(k\phi)$ for a fixed function $f$ and some transformation $k$ in $\phi$-space. Two variables that cannot be related in this way are said to be essentially different.

In [7], important elements of quantum mechanics are derived from such a situation assuming two related maximal accessible variables. The full derivation here relies on a group $K$ acting on $\phi$-space, and the concept of permissibility (; see Definition 1).

In [8,34] the following theorem is proved:
\bigskip

\textbf{Theorem 3} \textit{Assume that an observer Charlie has two maximal accessible related variables $\eta = f(\phi)$ and $\zeta = f(k\phi)$ in his mind. Then Charlie  cannot have in his mind another maximal accessible variable which is related to $\eta$ but essentially different from $\zeta$.}
\bigskip

Going back to the situation above, we then have: Look at particle pairs emerging in the vicinity of a black hole. Alice is all the time able to observe the particles that are escaping and leaving the region (the sources of Hawking radiation; see below), while Bob is only able to study the particles absorbed by the black hole. This is of course an ideal thought experiment, but much of the literature in this area is based on thought experiments.

 Assume further that the particle pairs are entangled with respect to the two properties $\eta$ and $\zeta$, where $\eta$ is a fixed function of (ideal) position $\theta$, while $\zeta$ is fixed function of (ideal) momentum $\xi$. Both Alice and Bob are interested in finding some measure of entropy related to their observations. As discussed in Section 6, entropy is closely related to Shannon information, and Shannon information may depend upon which variable we have a probability distribution over. So, by observing many particles, Alice can have two possible measures of entropy for her particles, one based upon the observations $\eta$, and another based on the complementary observations $\zeta$. Similarly, Bob has two possible measures of entropy for his particles, one based on his observations $\eta' = \eta'(\theta')$ on the absorbed particles, and one based upon his complementary observations $\zeta' = \zeta'(\xi')$.

From results of [7], the observer Alice can make a quantum model based on her two maximal variables $\theta$ and $\xi$ for her particles, while Bob can make a quantum model based on $\theta'$ and $\xi'$ for his particles. The predictions from these models may be sent to Charlie. From this, he has probability statements, based on quantum theory, both for $\theta'$ and for $\xi$, say, if these are the variables focused on by Alice and Bob, respectively, so from this, he can make a joint quantum model for all measured variables, and from this calculating the von Neuman entropy
\begin{equation}
S=-k_B\mathrm{trace}(\rho \mathrm{ln}(\rho)).
\label{xx}
\end{equation}

This formula depends crucially on how Charlie perceives the density matrix. It can be written as $\rho =\rho_A\otimes\rho_B$, where $\rho_A$ is the density matrix as perceived by Alice, and $\rho_B$ is the density matrix as perceived by Bob. 

Now go back to the choices that Alice and Bob had. We can imagine two different measure series for Alice, one where $\rho_A$ is based upon $\eta$, giving $\rho_{A,1}$, and one where $\rho_A$ is based upon $\zeta$, giving $\rho_{A,2}$. Similarly, Bob has two choices, and in summery, this gives 4 possible values for the entropy $S$. Charlie may want to report some weighted avarage over these 4 values. But this will require that he has a joint probability distribution over all four variables $\eta, \zeta, \eta'$ and $\zeta'$, but using Theorem 3 in the same way as it is used in [8,34], one can show that it is impossible to construct such a joint distribution. Hence, Charlie must make a subjective choice of what entropy to report.

\section{General relativity; Schwarzschield geometry}
\label{sec10}

For a spherical symmetrical system, like a star or a black hole, it is convenient to change from coordinates $ (t,x,y,z)$ to spherical coordinates $ (t,r,\theta, \phi)$, where $x=r\mathrm{sin}(\theta)\mathrm{cos}(\phi)$, $y = r\mathrm{sin}(\theta)\mathrm{sin}(\phi)$ and $z=r \mathrm{cos}(\theta)$. In the region outside the star/ black hole, one can then show [23] that the most general metric tensor depends on a mass $M$ and is given by
\[ g_{tt}=-(1-\frac{2GM}{cr}),\]
\[g_{rr}= (1-\frac{2GM}{cr})^{-1},\]
\[g_{\theta\theta}=r^2,\ \ \ \ \ \ \ \ \ \ \]
\[g_{\phi\phi}=r^2\mathrm{sin}^2(\theta),\]
and with all cross-terms vanishing. Here, $G$ is Newton's gravity constant, and $c$ is the velocity of light. This metric is called the Schwarzschield metric.

I will concentrate on the black hole case, where the metric has a singularity (the horizon) for $r=2GM/c$, and where $g_{tt}$ and $g_{rr}$ change sign in the interior $r<2GM/c$. In my terminology, the coordinates $(t,r,\theta,\phi)$ are inaccessible variables in the interior of a black hole. There is no mechanism by which these variables can be measured by an external observer. So I will concentrate on the outside region $r>2GM/c$, where the coordinates are accessible.

\section{On the theories of black holes}
\label{sec11}

An important new insight into a possible theory combining quantum mechanics came when Hawking [25] argued that black holes create and emit particles as they were hot bodies; see also the historical overview by Hawking and Isreal [26].

In [25] it is proposed that quantum mechanical effects cause black holes to create and emit particles as if they were hot bodies with temperature $\hbar\kappa/2\pi k_B$, where $\kappa$ is the surface gravity of the black hole and $k_B$ is Bolzmann's constant. The generalized entropy of the universe can, according to Hawking be taken as $S+\frac{k_B c^3}{4G\hbar}A$, where $S$ is the entropy outside black holes, $A$ is the sum of the surface areas of all black holes, and $G$ is Newton's gravity constant. The same formula can be used for the generalized entropy associated with a particular black hole, where $A$ now is the area of the horizon of that particular black hole, and $S$ is the entropy outside this black hole.

This generalized entropy never decreases. As a side remark, the fact that the entropy of a black hole is proportional to the area of its horizon, is related to the holographic principle proposed by t'Hooft and Susskind [27,28]: To decribe particle states in the vicinity of black holes, a two-dimensional function is required, the distribution over a two-dimensional coordinate on the horizon.

I will not here go into the details of black hole thermodynamics, which is reviewed in [29].

The entropy of Hawking radiation is discussed in detail in the recent article [30]. In this article, the Central Dogma of black holes is emphasized: \textit{As seen from the outside, a black hole can be described in terms of a quantum system with entropy $\frac{k_B c^3}{4G\hbar}A$ that evolves unitarily under time evolution.}

I will relate my treatment of black holes partly on the latest theoretical developments as they are discussed in [30] and in a  recent articles [31] in Scientific American. According to [31], both the socalled firewall paradox and the information paradox (black holes had seemed to contradict the basic physical principle that information is never lost) can be solved by considering a theory of wormholes: As a consequence of general relativity there is a non-vanishing probability that different black holes may be connected, a mechanism that was proposed already by Einstein and Rosen [32] in 1935.

I will have respect for these theories, but in a way they are speculations. Provided that these theories do not have consequences for accessible variables, they must be seen as having the same status as speculations around what happened `before' the Big Bang. In my view, theories involving just inaccessible variables should be avoided, and I will try to base my discussion upon information/ entropy related to accessible variables.

As described in Section 6, information and entropy are two sides of the same coin. And, as I see it, the amount of (Shannon) information in a physical system depends in a crucial way upon the observer(s) of the system.

Consider first the situation in Section 7 with two observers A and B moving with velocity $v$ with respect to each other. Suppose that A always satisfies the principle that information is never lost, and let A choose his information as based upon eiter the time-space vector $\theta$ or the energy-momentum vector $\xi$. Let B move very close to the black hole horizon, and let B be in the future light cone for A. Then the information of B is a function of the information of A, found by a suitable transformation as either information on $\theta'$ or $\xi'$. And the transformation here is invertible. Hence we conclude from this, that however close to the horizon B moves, he satisfies the same principle as A: information is never lost.

In the literature, information is also discussed for `observers' inside the black hole. According to my theory, this does not make sense: I would say that every variable that is connected to the inside of the black hole is inaccessible. This can be related to any theory. I base my approach to quantum theory closely on the distinction between accessible and inaccessible variables, and this should also be possible to translate to any quantummechanical effect caused by black holes. Using such a principle, some of the paradoxical discussions in the literature must be discarded, as I see it. 

The theory of Hawking radiation is based on entangled particle pairs arising spontanously from the vacuum. Now go back to the thought experiment of Section 9, where Alice observes one particle of such a pair, the particle that escapes as Hawking radiation, Bob observes the particle that is absorbed by the black hole, A and B are spacelikely separated, and both are observed by the actor Charlie. As argued in Section 9, Charlie must make a subjective choice of which entropy/ information to report. Assume that this choice has been made. The choice must be made independently for each particle pair, and in summary, Charlie must report an average or summed entropy. This total entropy must be independent of the subjective choices made by Charlie. By the theory above it must be proportional to the area of the event horizon.

By the law of large numbers this implies that the expectation $E(S)$, where $S$ is given by (\ref{xx}) must be independent of the choices of measurements that Alice and Bob have made. Here $\rho=\rho_A \otimes \rho_B$, so $\mathrm{ln}(\rho)=\mathrm{ln}(\rho_A)\mathrm{ln}(\rho_B)$, and it follows from (\ref{xx}) that
\begin{equation}
E(S)=E(-k_B[\rho_A\mathrm{ln}(\rho_A)]+ E(-k_B[\rho_B\mathrm{ln}(\rho_B)]=E(S_A)+E(S_B).
\label{xxx}
\end{equation}

This must be a constant $K$, which is independent of the choices made by Alice and Bob. Specifically, $K$ must be proportional to the surface area of the black hole's event horizon. Hence, we conclude from this that
\begin{equation}
E(S_B)=-E(S_A)+K.
\label{xxxx}
\end{equation}
The expected information that enters the black hole is a constant minus the expected information that escapes in the Hawking radiation. This conclusion is independent of any choice of measurement that the actors have made.

In this sense we have conservation of (expected) information. Note that this conclusion is made only by considering what happens outside the black hole. No strange hypoteses have been made on the inaccessible interior of the black hole, in particular the existence of any wormholes connected to this interior. I suggest that further theories of black holes also should have such a basis.

\section{Discussion}
\label{sec12}

The purpose of this paper has been to sketch a new, and in my opinion quite promising, attempt to understand parts of quantum theory and general relativity theory from a common basis. One important background for me is that relativity theory is developed by the mind of a single person, while quantum theory, in the way that it has existed up to now, is a patchwork of contributions from many different persons. Empirically, both theories have been verified to an impressive degree, but the foundation of and interpretation of quantum theory has been the source of much confusion. The book [5] is an attempt to develop the epistemic side of a new foundation, and from this propose a new interpretation: In every application, but also more generally, it is connected to the mind of a single actor or to the joint mind of a group of communicating actors.

One important background for the development of [5] has been that, in my opinion there was too little communication between researchers working with the foundation of quantum theory and researchers from different communities, say the statistics society. My book has been an attempt to develop elements of a future common culture.

What is culture? According to the author and philosopher Ralph D. Stacey it is a set of attitudes, opinions and convictions that a group of people share about how one should act towards each other, how things should be evaluated and done, which questions are important and which answers may be accepted. The most important elements in a culture are unconscious, and cannot be forced upon one from the outside.

One hope now is that results like Theorem 1 and Theorem 2 above, or similar results, in some future may be a part of a common culture among researchers in quantum foundation and in theoretical statistics. If this happens, I feel that it also should be easier to arrive at some joint understanding of physical theories describing the microscopic world and other  theories describing the macroscopic world. The foundation of quantum theory described here seems to be particularly relevant to such an understanding. This is further discussed elsewhere [5,6].

\section*{References}

[1] Rovelli, C. (2017). \textit{Reality is Not What it Seems.} Riverhead Books, New York.

[2] Susskind, L. and Lindsay, J. (2005). \textit{An Introduction to Black Holes, Information and the String Theory Revolution.} World Scientific, New Jersey.

[3] Laudal, O.A. (2021). \textit{Mathematical Models in Science.} World Scientific, New Jersey.

[4] Hardy, L. (2016). Operational general relativity: possibilistic, probabilistic, and quantum. arXiv: 1608.06940 [gr-qc].

[5] Helland, I.S. (2021) \textit{Epistemic Processes. A Basis for Statistics and Quantum Theory.} Second Edition. Springer, Berlin.

[6] Helland, I.S. (2023a). An alternative foundation of quantum mechanics. arXiv: 2305.06727 [quant-ph].

[7] Helland, I.S.(2010). \textit{Steps Towards a Unified Basis for Scientific Models and Methods.} Singapore: World Scientific.

[8] Helland, I.S. (2021a). The Bell experiment and the limitations of actors. Found. Phys. 52, 55.

[9] Helland, I.S.(2022b). On reconstructing parts of quantum theory from two related maximal conceptual variables. \textit{Intern. J. Theor. Phys.} \textbf{61}, 69. Correction. \textit{Intern. J. Theor. Phys.} \textbf{62}.

[10] Helland, I.S. (2023b). A simple quantum model linked to a theory of decisions. \textit{Foundations of Physics} 53, 12.

[11] Zwirn, H.: The measurement problem: Decoherence and convivial solipsism. Found. Phys. 46, 635-667. (2016).

[12] Schmid, D., Selby, J.H. and Spekkens, R.W. (2021). Unscrambling the omelette of causation and inference: The framework of causal-inferential theories. arXiv: 2009.03297 [quant-ph].

[13] Evans, P.W. (2021). The end of classical ontology for quantum mechanics?\textit{ Entropy} \textbf{23} (1), 12.

[14] Helland, I.S. (2021). Epistemological and ontological aspects of quantum theory. arXiv 2112.10484 [quant-ph].

[15] Casalbuoni, R. (2011). \textit{Introduction to Quantum Field Theory.} World Scientific, New Jersey.

[16] Shannon, C.E. (1948). The mathematical theory of communication. \textit{Bell System Technical Journal} \textbf{27}, 379-423, 623-656.

[17] Jaynes, E.T. (1957). Information theory and statistical mechanics. \textit{Physical Review} \textbf{106} (4), 620-630.

[18] Landauer, R. (1961). Irreversibility and heat generation in the computing process. \textit{IBM Journal of Research and Development} \textbf{5} (3), 183-191.

[19] Liang, J., Shi, Z. and Wierman, M.J. (2006). Information entropy, rough entropy and knowledge granulation in incomplete information systems. \textit{International Journal of General Systems} \textbf{35} (6), 641-654.

[20] Zurek, W.H. (1989). Algorithmic randomness and physical entropy. \textit{Physical Review A} \textbf{40} (8), 4731-4751.

[21] Gr\"{u}nwald, P. and Vit\'{a}nyi, P. (2004). Shannon information and Kolmogorov complexity. arXiv:cs/0410002 [cs.IT]

[22] Wigner, E. (1939). On unitary representations of the inhomogeneous Lorentz group. \textit{Annals of Mathematics} \textbf{40} (1), 149-204.

[23] Beskal, S., Kim, Y.S. and Noz, M.E. (2019). Poincar\'{e} symmetry from Heisenberg's uncertainty relations. \textit{Symmetry} \textbf{11}, 409-417.

[24] Schutz, B.F. (1990). \textit{ A First Course in General Relativity} Cambridge University Press, Cambridge.

[25] Hawking, S.W. (1975). Particle creation by black holes. \textit{Commun. Math. Phys,} \textbf{43}, 199-220.

[26] Hawking, S.W. and Israel, W. (1979). Introductory survey. In: Hawking, S.W. and Israel, W. [Ed.] General Relativity. An Einstein Centary Survey. Cambridge University Press, Cambridge.

[27] Susskind, L. (1994). The wold as a hologram. arXiv: 9409089 [hep-th].

[28] t'Hooft, G. (2000). The holographic principle. Opening lecture. arXiv: 0003004 [hep-th].

[29] Wall, A.C. (2018). A survey of black hole thermodynamics. arXiv: 1804.10610 [gr-qc].

[30] Almheiri, A., Hartman, T., Maldacena, J. Shaghoulian, E. and Amirhossein, T. (2021). The entropy of Hawking radiation. \textit{Rev. Mod. Phys.} \textbf{93} (3), 035002.

[31] Almheiri, A. (2022). Black holes, wormholes and entanglement. \textit{Scientific American} September 2022, 34-39.

[32] Einstein, A. and Rosen, N. (1935). The particle problem in the general theory of relativity. \textit{Phys. Rev.} \textbf{48}, 73.

[33] Helland, I.S. (2019). When is a set of questions to nature together with sharp answers to those questions in one-to-one correspondence with a set of quantum states? arXiv: 1909.08834 [quant-ph].

[34] Helland, I.S. (2023). On the Bell experiment and quantum foundation. arXiv: 2305.05299 [quant-ph].

\end{document}